\documentclass[%
 reprint,
 amsmath,amssymb,
 aps,
]{revtex4-2}

\usepackage{multirow}   
\usepackage{graphicx}
\usepackage{dcolumn}
\usepackage{bm}
\usepackage{tabularx}
\usepackage{booktabs}
\usepackage{float}
\makeatletter
\let\newfloat\newfloat@ltx
\makeatother
\usepackage{caption}
\usepackage{ragged2e}
\usepackage{hyperref}

\DeclareCaptionFormat{myformat}{\justifying #1#2#3}
\usepackage{xcolor,colortbl}
\usepackage[table]{xcolor}
\usepackage{algorithm}
\usepackage[english]{babel}
\usepackage{algpseudocode}
\usepackage{adjustbox}
\usepackage{makecell}
\usepackage{amsmath}

\usepackage{nicematrix}
\usepackage{adjustbox}
\usepackage{multirow}
\usepackage{makecell}
\usepackage[table]{xcolor}
\newcolumntype{Y}{>{\centering\arraybackslash}X}

\begin{document}

\preprint{APS/123-QED}

\title{Constraint-Optimal Driven Allocation for Scalable QEC Decoder Scheduling}

\author{Dongmin Kim, Jeonggeun Seo, and Youngsun Han}
 \thanks{youngsun@pknu.ac.kr}%
\affiliation{%
Department of AI Convergence, Pukyong National University, Busan 48513, South Korea
}%
\author{Yongtae Kim}
\affiliation{%
School of Computer Science and Engineering, Kyungpook National University, Daegu 41566, South Korea
}%

\date{\today}

\begin{abstract}

Fault-tolerant quantum computing (FTQC) requires fast and accurate decoding of Quantum Error Correction (QEC) syndromes. However, in large-scale systems, the number of available decoders is much smaller than the number of logical qubits, leading to a fundamental resource shortage. To address this limitation, Virtualized Quantum Decoder (VQD) architectures have been proposed to share a limited pool of decoders across multiple qubits. While the Minimize Longest Undecoded Sequence (MLS) heuristic has been introduced as an effective scheduling policy within the VQD framework, its locally greedy decision-making structure limits its ability to consider global circuit structure, causing inefficiencies in resource balancing and limited scalability. In this work, we propose Constraint-Optimal Driven Allocation (CODA), an optimization-based scheduling algorithm that leverages global circuit structure to minimize the longest undecoded sequence length. Across 19 benchmark circuits, CODA achieves an average 74\% reduction in the longest undecoded sequence length. Crucially, while the theoretical search space scales exponentially with circuit size, CODA effectively bypasses this combinatorial explosion. Our evaluation confirms that the scheduling time scales linearly with the number of qubits, determined by physical resource constraints rather than the combinatorial search space, ensuring robust scalability for large-scale FTQC systems. These results demonstrate that CODA provides a global optimization-based, scalable scheduling solution that enables efficient decoder virtualization in large-scale FTQC systems.

\end{abstract}

\maketitle

\section{Introduction}

Quantum computing has attracted significant attention from both academia and industry due to its potential to provide computational advantages in problems such as integer factorization~\cite{shor1994algorithms, RevModPhys.68.733}, unstructured search~\cite{grover1996fast,10.5555/870802}, and complex quantum simulations~\cite{lloyd1996universal,RevModPhys.86.153}. Recent advances in superconducting, trapped-ion, and photonic hardware platforms, along with increasingly sophisticated fabrication techniques, have enabled the execution of small to medium-scale quantum circuits~\cite{arute_quantum_2019,bruzewicz_trapped-ion_2019,zhong_quantum_2020,m2020,monroe2013}. However, due to the intrinsic fragility of quantum states, these systems remain fundamentally vulnerable to noise and errors, and are therefore referred to as belonging to the Noisy Intermediate-Scale Quantum (NISQ) era~\cite{preskill2018quantum, arute2019quantum}. To achieve practical quantum advantage, it is essential to scale beyond NISQ devices toward large-scale Fault-Tolerant Quantum Computing (FTQC) systems with thousands of high-fidelity qubits and deep circuit depths~\cite{fowler2012surface, campbell2017roads,litinski2019}.


As FTQC systems scale, the increasing number of qubits and circuit depth amplify the impact of noise and hardware imperfections, leading to rapid error accumulation~\cite{gambetta2017building, krinner2022realizing}. Specifically, decoherence, gate infidelity, and crosstalk fundamentally limit the reliability and scalability of current quantum processors~\cite{sarovar2020detecting, krinner2022realizing}. To mitigate these issues, FTQC employs Quantum Error Correction (QEC), which encodes logical information into multiple physical qubits and continuously measures error syndromes to suppress logical error rates below a target threshold~\cite{gottesman1997stabilizer, fowler2012surface}. The task of interpreting these error syndromes and determining appropriate correction operations is carried out by the QEC decoder. Depending on the underlying decoding algorithm, the QEC decoder serves as a critical component that directly affects the performance and stability of the entire FTQC system~\cite{fowler2012surface, dennis2002topological, chamberland2020building,gqai2025naturebelowthr}.


In large-scale FTQC systems, decoders which is a critical component that determines error correction performance, pose a major scalability constraint on the expansion of large-scale quantum systems. This is because physical limitations in power, memory bandwidth, interconnect, and chip area make it practically infeasible to assign a dedicated decoder to each logical qubit~\cite{campbell2017roads, das2024scalable,coms2021,Camps2024}. As a result, the number of available decoders is smaller than the number of logical qubits, leading to a structural resource imbalance that serves as a fundamental bottleneck in the scalability of FTQC systems. To address this imbalance, limited decoder resources inevitably be shared among multiple qubits over time.


Prior work proposed an architectural approach based on Virtualized Quantum Decoder (VQD) architectures to overcome the imbalance problem~\cite{das2024scalable, chamberland2020building,maurya2024classicalqec}. VQD combines a small number of decoders into a shared pool and allows multiple logical qubits to use them in a time-multiplexed manner, thereby improving the efficiency of limited decoder resources usage. This concept is analogous to time-multiplexing in classical computing systems~\cite{10.1145/321738.321743}. In addition to the VQD architecture, several scheduling algorithms have been proposed, including Most-Frequent Decoder (MFD), Round-Robin (RR), and Minimize Longest Undecoded Sequence (MLS)~\cite{das2024scalable}. These algorithms are all based on local heuristics and therefore fail to anticipate future decoding demands required by the global circuit structure. Among them, MLS provides good short-term performance by prioritizing the qubits that have accumulated the longest backlog of undecoded syndromes (i.e., the longest undecoded sequence), but its reliance on local heuristics prevents it from considering global workload distribution and achieving an optimal solution. In addition, the extra computational overhead required for calculating priorities introduces scalability limitations that become increasingly exacerbated as circuit size grows.

This limitation becomes particularly important in circuits containing non-Clifford operations, especially logical $T$-gate execution~\cite{bravyi_universal_2005}. In many fault-tolerant schemes, logical $T$-gates are realized via magic-state-based gate teleportation. While fault-tolerant computation can proceed in a Pauli or Clifford frame under slow error diagnostics~\cite{chamberland_fault-tolerant_2018}, logical $T$-gate teleportation requires the logical Z measurement to be decoded before the classically controlled S correction is applied~\cite{skoric_parallel_2023}. Therefore, the decoded outcome needs to be available when logical $T$-gate execution proceeds, which in practice means that decoding is completed immediately beforehand. MLS makes allocation decisions myopically based only on the current backlog. This prevents it from anticipating how future $T$-gate-related decoding demands constrain decoder availability for the remaining qubits. As a result, decoder resources may be repeatedly consumed by those upcoming demands, while other qubits accumulate unexpectedly long undecoded sequences.


To overcome the limitations of existing scheduling algorithms, we propose Constraint-Optimal Driven Allocation (CODA), a global optimization-based decoder scheduling algorithm for FTQC systems. Specifically, the CODA minimizes the longest undecoded sequence length across all qubits by jointly considering global circuit structure, which captures the temporal and spatial distribution of decoder requests, including future mandatory decode demands, and hardware resource constraints. This approach achieves shorter longest undecoded sequence lengths by balancing decoder resource usage, while ensuring better scalability compared to existing heuristic approaches.


We integrate the CODA algorithm into an existing Python-based decoder scheduling simulator and evaluate it alongside RR and MLS under identical conditions. Across 19 benchmarks comprising representative quantum programs, CODA achieves a 74\% reduction on average in the longest undecoded sequence length and demonstrates robust scalability by overcoming theoretical exponential complexity, ensuring practical scheduling times even at larger circuit scales. These results demonstrate that CODA goes beyond the performance limitations of existing heuristics and provides a scalable, global optimization-based scheduling solution under resource constraints, effectively enhancing the stability and scalability of FTQC systems.


The main contributions of this paper are summarized as follows:

\begin{itemize}
    \item \textbf{Mathematical modeling of scalable decoding:} 
    We analyze the computational hardness of the decoder scheduling problem and reformulate it into a sequence of feasibility decision problems to enable efficient solution finding.
    
    \item \textbf{CODA algorithm:} 
    We design a constraint optimization–based scheduling algorithm that incorporates resource and decoding priority constraints, and provides a scalable, global optimization-based scheduling solution.

    \item \textbf{Demonstration validation:} 
    We demonstrate that CODA significantly reduces the longest undecoded sequence lengths compared to state-of-the-art heuristics while maintaining linear scalability with respect to circuit size.
\end{itemize}

The rest of this paper is organized as follows. Section~\ref{sec:Background and Related Works} reviews the background of FTQC, QEC decoding, the scalability challenges, and the motivation of this work.
Section~\ref{sec:The CODA System Architecture} presents the overall process of the CODA scheduler and describes its algorithm.  
Section~\ref{sec:Result} describes the demonstration setup and presents comparative results against RR and MLS.
Finally, Section~\ref{sec:Conclusion} concludes the paper. Additional details and extended discussions are provided in the Appendix.


\section{\label{sec:Background and Related Works}Background}
In this section, we provide the technical background on FTQC, specifically focusing on the critical role of QEC decoding.
We then analyze the scalability constraints imposed by limited decoder resources in large-scale systems, identifying the fundamental bottlenecks that motivate our proposed constraint-optimal scheduling approach.


\subsection{FTQC and QEC Decoding}

Realizing large-scale FTQC requires low-latency detection and correction of physical errors that occur during computation through quantum error correction (QEC)~\cite{shor1995scheme,fowler2012surface,preskill2018quantum}, essential to prevent the accumulation of physical errors from causing logical failure.
QEC encodes a single logical qubit into multiple physical qubits with redundant encoding so that logical information can be preserved even if some physical errors occur.
Because the quantum state cannot be directly measured without destroying the encoded information, a set of commuting operators called stabilizers is periodically measured, and error-related information, referred to as syndromes, is extracted indirectly~\cite{gottesman1997stabilizer,dennis2002topological,google2021}.
Syndromes provide information about the presence and pattern of errors
without revealing the underlying logical state.

QEC operates as a repetitive cycle comprising stabilizer measurement, syndrome interpretation, and recovery operations. Within this cycle, the decoder serves a critical function by inferring error configurations using representative algorithms such as Minimum-Weight Perfect Matching (MWPM), Union-Find, and data-driven or machine learning–based methods to determine the necessary corrections~\cite{fowler2012surface,wang2010threshold,hastings2021decoding,delfosse2021almost}. This decoding process is computationally intensive and operates under strict latency constraints, directly determining the overall stability and reliability of FTQC systems. Given this criticality, realizing large-scale FTQC with a massive number of qubits ideally necessitates sufficiently abundant decoding resources to support the entire qubit array. However, as systems scale toward large-scale FTQC, the limited scalability of decoders emerges as a fundamental bottleneck, leading to the rapid accumulation of undecoded syndromes and the inability to apply timely corrections, thereby threatening the feasibility of scalable error correction.~\cite{fowler2012surface,das2024scalable,battistel2023}.

\subsection{\label{sec:Scalability Challenge and Motivation}Scalability Challenge and Motivation}

As large-scale FTQC systems scale to thousands or even millions of logical qubits, the number of available decoders does not increase proportionally~\cite{Jouppi2017,krste2006}. Under this scaling trend, a one-to-one decoder-to-qubit assignment becomes infeasible, resulting in a fundamental resource imbalance. This limits the ability to allocate decoders promptly when error syndromes occur, leaving some syndromes unprocessed in the queue. Since decoders are responsible for interpreting measured syndromes and applying corrections promptly, this shortage leads to the rapid accumulation of undecoded syndromes. This growing backlog makes the entire system vulnerable to logical failures due to the inability to decode errors in time, while simultaneously increasing the memory overhead required to store undecoded syndrome data. A conceptual illustration and a more detailed analysis of these scalability constraints are provided in Appendix~\ref{sec:Scalability Limits of QEC Decoders in FTQC}.

\begin{figure}[!t]

    \centering{\includegraphics[width=0.455\textwidth]{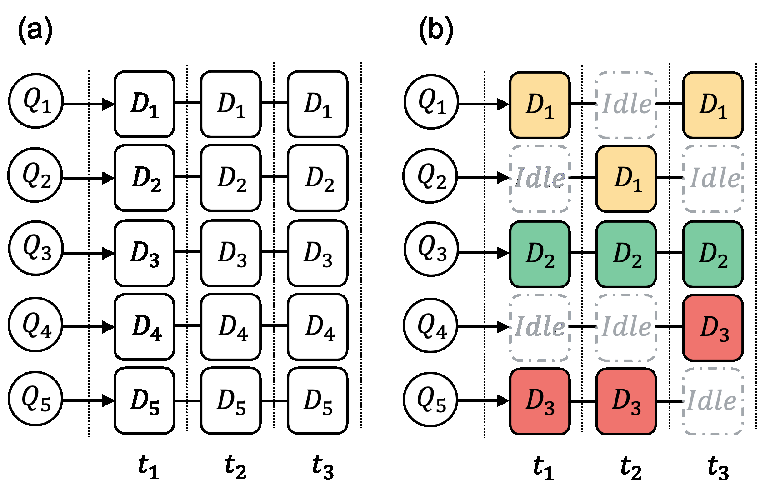}}
    \caption{Examples of decoder scheduling under different scenarios. (a) Ideal case, where the number of decoders matches the number of logical qubits. Each logical qubit $Q_i$ is continuously assigned to its dedicated decoder $D_i$ across all time slices, resulting in no idle states and no accumulation of undecoded data.
(b) Resource-limited case, where the number of available decoders is smaller than the number of logical qubits. Some qubits remain idle in certain time slices, and the same decoder is reused across multiple qubits over time.}
    \label{fig:Scheduling example}
\end{figure}

The VQD architecture is a structural framework that supports a larger number of logical qubits with a limited number of decoders. The core idea of VQD is that a single decoder can sequentially process different qubits across multiple time slices~\cite{paul2003,liu1973,mijumbi2016network}. Here, a time slice refers to a defined time unit in which decoding requests are generated and processed. This allows each decoder to handle one qubit at a time while covering multiple qubits over time, thereby enabling more logical qubits to be supported with fewer decoder resources. When the number of qubits is small, a one-to-one mapping between decoders and qubits is possible, as shown in Fig.~\ref{fig:Scheduling example}(a). However, as the system scales toward FTQC, the number of qubits increases, requiring an efficient allocation of limited decoders across multiple qubits over time, as illustrated in Fig.~\ref{fig:Scheduling example}(b). This scenario intuitively demonstrates the core operational principle of VQD, where the way limited decoders are scheduled plays a crucial role in determining the overall stability and performance of the QEC system. A more detailed description of the VQD structure is provided in Appendix~\ref{sec:Virtualization of Quantum Decoder Architecture}.


With the VQD structure, several QEC decoder scheduling policies have been proposed. As discussed in Appendix~\ref{sec:Virtualization of Quantum Decoder Architecture}, Maurya and Tannu introduced heuristic-based scheduling policies, including Most-Frequent Decoder (MFD), Round-Robin (RR), and Minimize Longest Undecoded Sequence (MLS). These heuristic approaches are simple to implement and achieve reasonable performance for small- to medium-scale circuits. However, they rely on local decision-making and fail to capture global dependencies in the circuit execution flow, leading to backlog accumulation as the system scales. Among them, MLS is considered the most effective heuristic; however, it faces two fundamental limitations as the system scales. First, its greedy allocation strategy fails to account for global circuit structure, specifically ignoring future mandatory decode demands associated with T-gate operations, which results in the accumulation of unexpected backlogs. Second, the computational overhead required for calculating priorities based on undecoded sequence lengths at every time step grows rapidly with circuit width and depth, imposing severe scalability constraints.


\begin{figure*}[!t] 
    \centering
    \centering{\includegraphics[width=17.5cm]{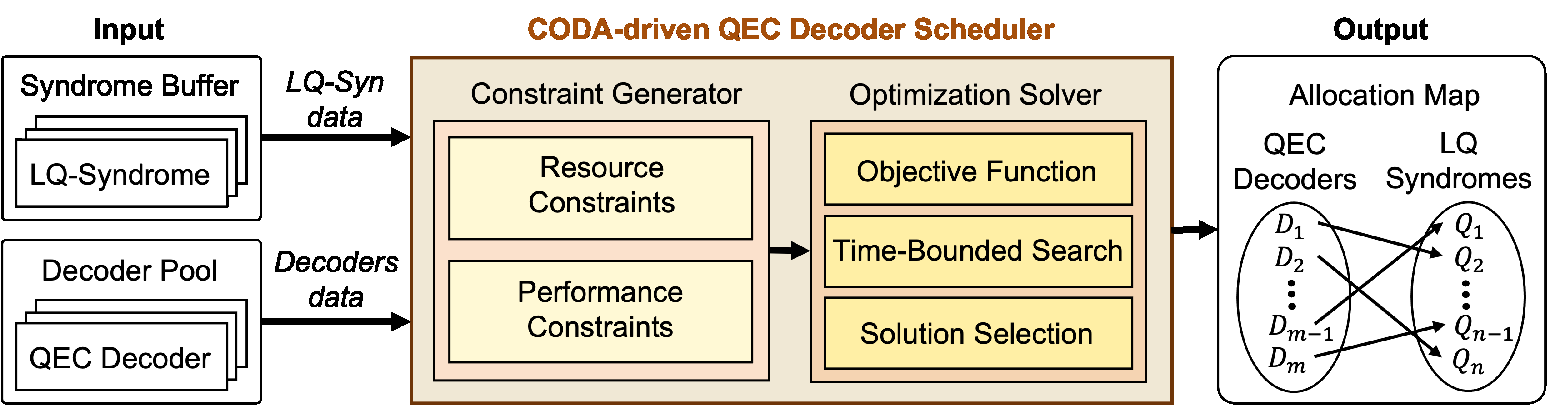}}
    \caption{Overall CODA-driven QEC decoder scheduler workflow. It takes two inputs: syndrome data from the syndrome buffer and decoder data from the decoder pool. The scheduler consists of two main components: (1) Constraint Generator that defines resource constraints (the number of available decoders) and performance constraints (allocatable position), and (2) Optimization Solver that determines the optimal schedule via the objective function (minimizing undecoded sequence length), time-bounded search (limiting computation time), and solution selection (identifying the minimum feasible gap value). After a series of processes, the scheduler returns the map that assigns decoders to logical qubits.}
    \label{fig:CODA Process}
\end{figure*}

The inherent limitations of heuristic-based scheduling policies and scalability constraints reveal that existing approaches are insufficient to support large-scale FTQC environments effectively~\cite{knuth1998art3}. In such environments, a new strategy is required to utilize limited decoder resources efficiently, minimize the accumulation of undecoded sequences, and fundamentally address the scalability challenges of decoder scheduling. To achieve this, we propose Constraint-Optimal Driven Allocation (CODA), which formulates the decoder allocation problem as a unified constraint optimization problem and employs a constraint programming (CP) solver to search for a globally optimized schedule, thereby overcoming the limitations of heuristic policies and enabling stable error correction under limited decoder resources~\cite{marriott1998guide}.

\section{\label{sec:The CODA System Architecture}Proposed Methodology}
In this section, we first present the overall system structure of the CODA-driven QEC decoder scheduler to illustrate the workflow of the CODA algorithm conceptually. We then present the detailed procedure of the CODA algorithm designed for efficient decoder scheduling. Finally, we clearly demonstrate the differences and advantages of the proposed approach over conventional scheduling methods through a scheduling example.

\subsection{Workflow of the CODA Scheduler}
Figure~\ref{fig:CODA Process} illustrates the overall execution workflow of the proposed CODA-driven QEC decoder scheduler. This structural overview provides a conceptual flow for understanding the core algorithm detailed in the Section~\ref{sec:CODA Algorithm}. The process operates by taking two specific inputs: syndrome data from the syndrome buffer and the current state of decoder resources in the decoder pool. Based on these inputs, it generates an optimal Allocation Map that efficiently distributes limited decoders among multiple logical qubits. The execution flow is organized into two consecutive phases: constraint formulation and optimization execution through Constraint Generator and Optimization Solver, respectively.


In the first phase, the Constraint Generator translates the physical status of resources into constraints required for the algorithm. This component defines the valid search space by establishing two categories of constraints. Resource Constraints reflect the aggregate capacity of $m$ available decoders, ensuring that simultaneous allocations across $n$ logical qubits do not exceed this physical limit. Simultaneously, Performance Constraints impose limits on the the longest length of undecoded sequences to prevent excessive backlog accumulation.

The subsequent phase is carried out by the Optimization Solver, which serves as the computational core where the actual allocation is determined based on the defined constraints. This component performs optimization through three key mechanisms. First, the Objective Function quantitatively defines the scheduling goals: minimizing the longest undecoded sequence length. Second, the Time-Bounded Search mechanism strictly limits the computation time to control complexity. Enforcing a strict upper bound on the search duration, it prevents the exhaustive search of all possible scheduling scenarios, thereby guaranteeing a predictable execution time. Third, the Solution Selection process utilizes a gap-incremental search strategy. By progressively increasing the gap parameter $G$, which represents the maximum allowable undecoded sequence length (starting from $G=1$), the solver identifies the minimum feasible gap value within the time limit and finalizes the corresponding solution as the final allocation map. The final output, the Allocation Map, represents the final allocation result derived from this procedure.

\subsection{Constraint-Optimal Driven Scheduling (CODA) Algorithm}
\label{sec:CODA Algorithm}

The fundamental computational difficulty of the decoder scheduling problem stems from the structural characteristic that allocation decisions at the current time slice directly determine the backlog state in subsequent time slices. Undecoded syndromes at time $t$ do not disappear but accumulate and carry over to the next time slice. As this phenomenon repeats, the total number of scheduling scenarios to explore explodes exponentially with circuit depth ($\Omega(\binom{N}{M}^T)$). Consequently, finding a mathematically perfect global optimum via exhaustive search is computationally intractable (NP-hard) for large-scale circuits, leading to exponential scheduling time (see Appendix~\ref{sec:Problem Formulation} for formal proof). To overcome these computational limitations, CODA alters the approach from solving the complex optimization problem directly. Instead of attempting to directly minimize the longest undecoded sequence length, the problem is simplified into a concrete verification process that determines whether a schedule satisfying all constraints exists within a given undecoded sequence length limit $G$. This transformation converts the uncertain search for an optimum into a stepwise verification procedure for a fixed target, thereby effectively controlling the computational complexity.

Algorithm~\ref{alg:CODA_Scheduling_Algorithm} details the specific procedure for executing this strategy, and the principal notation used in this algorithm and the accompanying discussion are summarized in Table~\ref{tab:main_notation}. The algorithm starts from the theoretically smallest possible limit of 1 and sequentially increments this limit $G$ by 1 until a feasible schedule is identified. In each iteration, the algorithm first initializes a mathematical model $\mathcal{M}$ for the current $G$ and defines decision variables, including the binary allocation variable $x_{d,q,t}$ and the integer backlog state variable $U_{q}(t)$. To define the feasible search space, three mandatory sets of constraints are injected into the model (Appendix~\ref{sec:Constraints}).


\begin{table}
\caption{Notation summary}
\label{tab:main_notation}
\centering
\renewcommand{\arraystretch}{1.08}
\setlength{\tabcolsep}{5pt}

\newcommand{\tabdef}[1]{%
  \parbox[t]{0.72\columnwidth}{%
    \justifying
    \setlength{\parindent}{0pt}%
    \strut #1\strut
  }%
}

\begin{tabular*}{\columnwidth}{@{\extracolsep{\fill}}ll}
\hline
\textbf{Symbol} & \textbf{Definition} \\
\hline
$S$ & \tabdef{Syndrome-derived scheduling input used by CODA, representing time-sliced decoding requests} \\[3pt]
$D$ & \tabdef{Set of available decoders} \\[3pt]
$L$ & \tabdef{Total number of time slices} \\[3pt]
$T_{\mathrm{limit}}$ & \tabdef{Solver time limit for each feasibility check} \\[3pt]
$A^*$ & \tabdef{Final allocation map selected by CODA} \\[3pt]
$G$ & \tabdef{Candidate upper bound on the longest undecoded sequence length} \\[3pt]
$d, q, t, \tau$ & \tabdef{Indices for decoder, logical qubit, current time slice, and T-gate time slice, respectively} \\[3pt]
$x_{d,q,t}$ & \tabdef{Binary assignment variable; $x_{d,q,t}=1$ if decoder $d$ is assigned to logical qubit $q$ at time slice $t$} \\[3pt]
$y_{q,t}$ & \tabdef{Decoding indicator; $y_{q,t}=1$ if logical qubit $q$ is decoded at time slice $t$} \\[3pt]
$U_q(t)$ & \tabdef{Backlog length of logical qubit $q$ at time slice $t$} \\[3pt]
$\mathcal{M}$ & \tabdef{Constraint model used in the feasibility check for a given $G$} \\[3pt]
$a_{d,t}$ & \tabdef{Decoder availability indicator; $a_{d,t}=1$ if decoder $d$ is available at time slice $t$} \\[3pt]
$T_\tau$ & \tabdef{Set of qubits scheduled to execute a T gate at time slice $\tau$} \\[3pt]
$A_{\mathrm{sol}}$ & \tabdef{Feasible allocation returned by CP-SAT for the current $G$} \\[3pt]
\hline
\end{tabular*}
\end{table}

First, resource constraints are applied to enforce physical hardware limits, ensuring that the number of qubits processed by a decoder does not exceed its availability and preventing duplicate assignments to a single qubit. Second, backlog evolution and bound constraints are established. The model explicitly defines the recurrence relation for backlog accumulation, in which unserved qubits increment their backlog, and imposes an inequality condition ensuring that $U_{q}(t)$ does not exceed the current limit $G$. Third, hard precedence constraints are applied to guarantee logical correctness. For qubits scheduled for T-gate operations (denoted as set $\mathcal{T}_{\tau}$), the condition that decoding must be completed immediately prior to gate execution ($y=1$) is strictly enforced to ensure circuit integrity.

\begin{algorithm}[bt!]
\caption{CODA Scheduling Algorithm}
\label{alg:CODA_Scheduling_Algorithm}
\begin{algorithmic}[1]
\Procedure{CODA}{$S, D, L, T_{limit}$}
    \State $A^* \leftarrow \emptyset$
    
    \For{$G = 1$ \textbf{to} $L$}
        \State \Comment{Define Decision Variables}
        \State $x_{d,q,t}, y_{q,t}, U_{q}(t) \leftarrow \text{InitializeVariables}(S, D, L)$
        \State \Comment{Construct Constraint Set $\mathcal{M}$}
        \State $\mathcal{M} \leftarrow \emptyset$
        \State $\mathcal{M} \leftarrow \mathcal{M} \cup \{ \sum_{q} x_{d,q,t} \le a_{d,t}, \sum_{d} x_{d,q,t} \le 1 \}$
        \State $\mathcal{M} \leftarrow \mathcal{M} \cup \{ U_q(t+1) = (1-y_{q,t})(U_q(t)+1) \}$
        \State $\mathcal{M} \leftarrow \mathcal{M} \cup \{ U_q(t) \le G \}$
        \State $\mathcal{M} \leftarrow \mathcal{M} \cup \{ y_{q,\tau-1} = 1 \mid \forall \tau \in T, \forall q \in \mathcal{T}_{\tau} \}$
        \State \Comment{Solve Optimization Problem}
        \State $A_{sol} \leftarrow \text{CP-SAT}(\mathcal{M}, T_{limit})$
        
        \If{$A_{sol} \neq \emptyset$}
            \State $A^* \leftarrow A_{sol}$
            \State \textbf{break}
        \EndIf
    \EndFor
    
    \State \textbf{return} $A^*$
\EndProcedure
\end{algorithmic}
\end{algorithm}

\begin{figure*}[!t] 
    \centering
    \centering{\includegraphics[width=13.5cm]{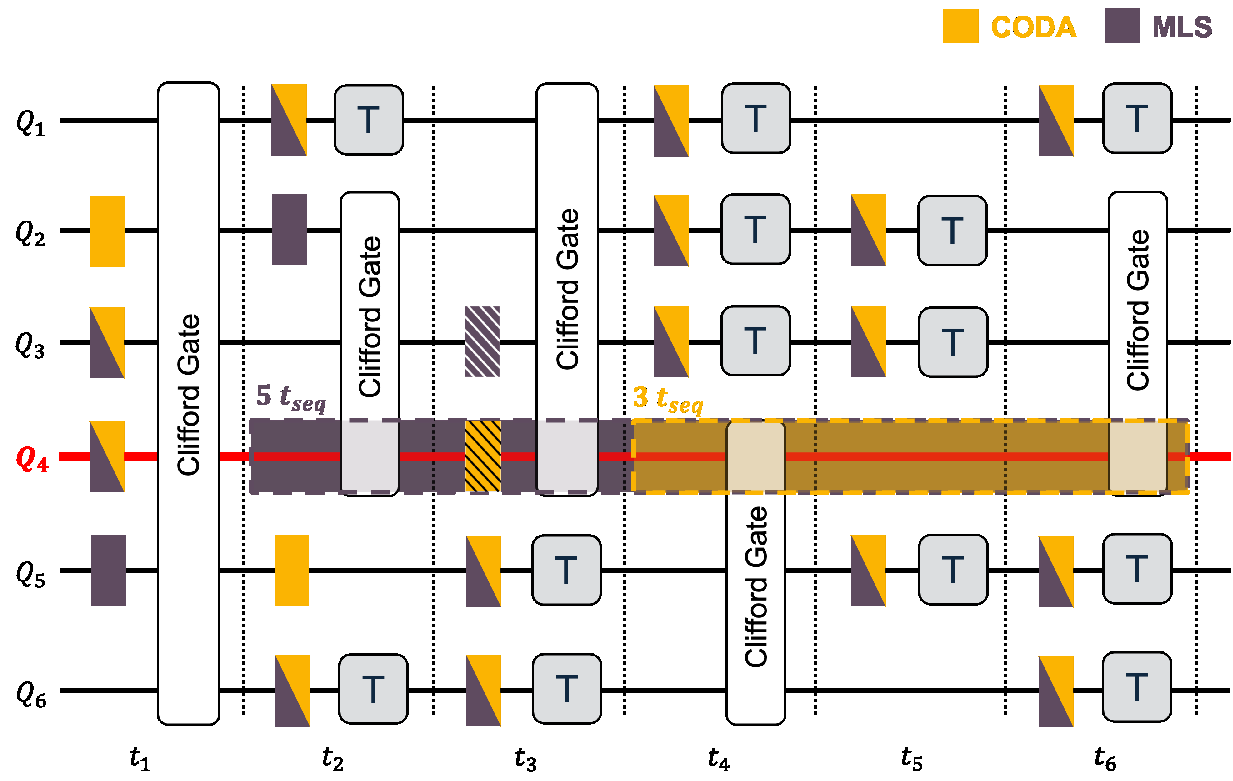}}
    \caption{Scheduling example comparing CODA with MLS for 6 logical qubits and 3 available decoders. MLS prioritizes the qubit with the longest undecoded sequence at each time slice but cannot anticipate mandatory decoder allocations required by future $T$-gate operations, leading to extended undecoded sequences such as $5t_{seq}$ for $Q_4$. In contrast, CODA reduces the longest undecoded sequence length to $3t_{seq}$ by employing a global optimization approach that accounts for future precedence constraints. The hatched decoding boxes at $t_3$ highlight the critical decision point between the two policies: anticipating the upcoming $T$-gate demand on $Q_3$, CODA allocates the decoder to $Q_4$ instead, whereas MLS does not.}
    \label{fig:CODA and MLS}
\end{figure*}

The fully formulated model is passed to the CP-SAT solver along with a predefined time limit ($T_{limit}$). The solver explores whether a solution satisfying all constraints exists within the limited time, based on the objective function of minimizing the longest undecoded sequence length (Appendix~\ref{sec:Optimization Strategy}). Since the algorithm searches sequentially from the smallest $G$, finding a feasible solution logically proves that the current $G$ is the minimum achievable bound. Therefore, if a solution is found, the algorithm immediately terminates the search and returns the allocation map. On the other hand, if a solution is not found within the time limit, the algorithm determines that the current conditions are too strict, increments $G$ by 1 to relax the constraints, and proceeds to the next search step (Appendix~\ref{sec:Solver Configuration}).

This time-bounded search mechanism bypasses exponential computational complexity and ensures linear scalability. The total execution time is proportional to the product of the resulting gap value $G_{final}$ and the time limit $T_{limit}$ of each step. In particular, since the minimum required time is determined by the ratio of workload (number of qubits) to processing resources (number of decoders), the value of $G_{final}$ tends to increase linearly in proportion to the number of qubits. A detailed analysis of this scaling behavior is provided in Appendix~\ref{sec:Complexity}. Furthermore, as $G$ increases, the constraints are relaxed, and the region where solutions can exist expands exponentially. Thus, even in worst-case scenarios characterized by a highly limited number of decoders relative to a large number of logical qubits, the exponential expansion of the feasible region ensures rapid algorithmic convergence. The mathematical proof of this convergence is provided in Appendix~\ref{sec:Complexity}. This guarantees that CODA can reliably identify a globally optimized schedule in large-scale circuits without exhaustively examining all possibilities.

\subsection{Scheduling Example}


Figure~\ref{fig:CODA and MLS} illustrates the difference between MLS and CODA scheduling. 
MLS assigns decoders to the qubit with the longest undecoded sequence at each time slice. 
While this policy is intuitive and may appear locally efficient, it fails to anticipate mandatory decoder allocations required by future $T$-gates in subsequent time slices. Specifically, if the number of qubits requiring mandatory decoding at a specific time slice consumes all available decoders, qubits with accumulated backlogs from previous time slices are inevitably deprived of decoding support.
Consequently, when decoders are preemptively consumed by $T$-gate operations, the remaining qubits may continue without decoding support for several time slices, leading to excessive backlog accumulation. In the illustrated example, qubit $Q_4$ experiences this limitation, resulting in the longest undecoded sequence length of $5t_{\mathrm{seq}}$ under MLS. This phenomenon becomes increasingly severe in circuits with a high density of $T$-gates relative to the limited number of decoders, exacerbating worst-case backlogs.

In contrast, CODA adopts a global optimization approach that iteratively evaluates the entire circuit under progressively increasing limit $G$ on the longest undecoded sequence length. CODA treats $G$ as a feasibility threshold and iteratively tests values of $G$ (starting from $G=1$) to identify the minimum $G$ for which a feasible schedule exists. In other words, CODA does not merely respond to the longest undecoded sequence at the current time slice. Instead, it performs a circuit-wide optimization that anticipates future $T$-gate demands and adjusts decoder allocations across all logical qubits. Through this global perspective, CODA prevents localized bottlenecks and achieves globally balanced decoder resource usage, ensuring fair access to resources for all qubits.
As a result, in the same scenario, qubit $Q_4$ receives a decoder earlier, reducing its longest undecoded sequence length to $3t_{\mathrm{seq}}$. This example demonstrates that CODA provides a more stable and balanced resource allocation than MLS, effectively mitigating worst-case backlogs even under constrained decoder resources.

\section{\label{sec:Result}Result}
This section presents the results of numerical demonstrations of the proposed CODA scheduling policy. We evaluate its performance through simulations on various benchmark circuits and comparative scheduling policies, providing a quantitative analysis of CODA’s effectiveness and a discussion of its advantages over existing approaches.

\begin{figure*}[!t] 
    \centering
    \centering{\includegraphics[width=17.5cm]{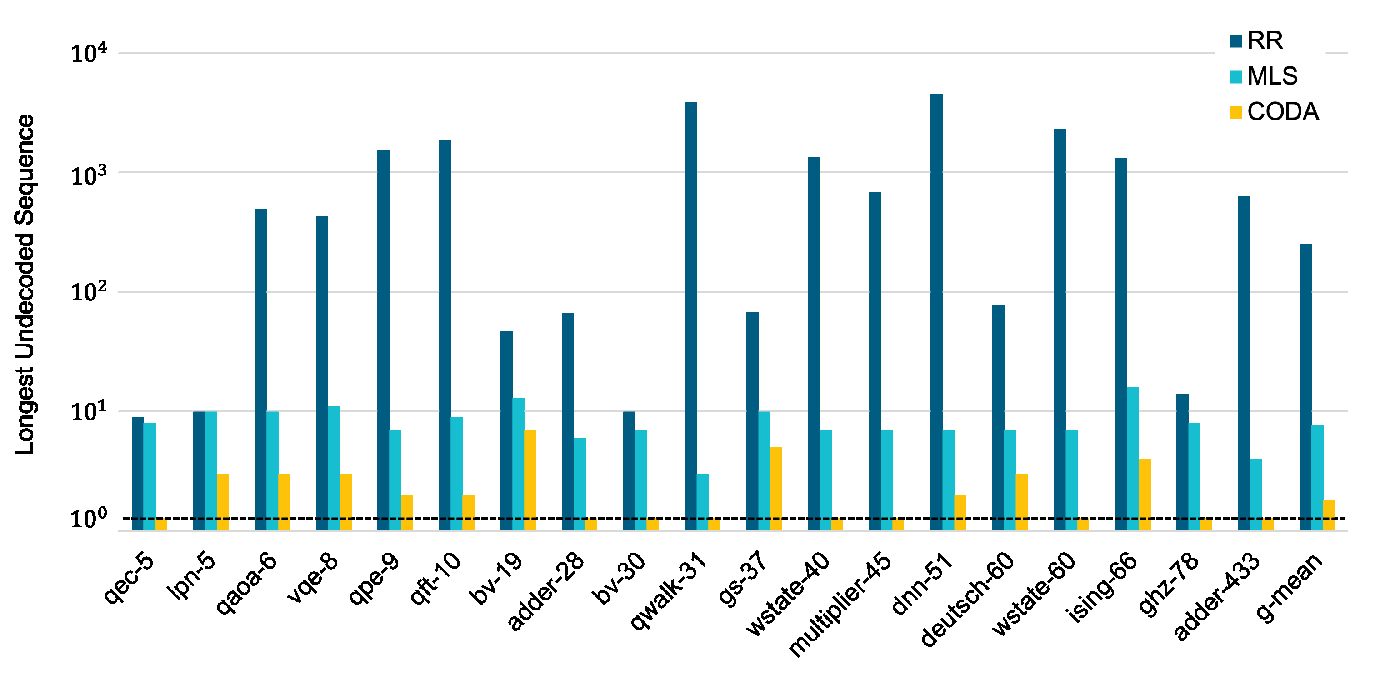}}
    \caption{Comparison of the longest undecoded sequence length obtained from RR, MLS, and the proposed CODA scheduling policy over a range of benchmark circuits, as well as their geometric mean (g-mean). The results demonstrate that CODA consistently shows the shortest undecoded sequences among all benchmarks, with precise numerical values reported in Table~\ref{tab:benchmark_results}.}
    \label{fig:Longest Undecoded Sequence Comparison}
\end{figure*}

\subsection{Demonstration Setup}

To evaluate the performance of the proposed CODA scheduling policy, we integrated the CODA algorithm into the Python-based decoder allocation simulator provided within the VQD framework~\cite{noauthor_anonymized_nodate}. To generate the input required for this simulator, we employed the Lattice Surgery Compiler to convert various quantum circuits into the Lattice Surgery Intermediate Representation (LLI) format~\cite{LSC}. The simulator then parses this LLI output to divide the circuit execution into discrete time slices, while tracking quantum operations and decoding requests at each slice to generate the scheduling workload for CODA.
All demonstrations were implemented in Python 3.11, and constraint optimization was performed using the OR-Tools CP-SAT solver~\cite{ortools_cp_sat}. The simulations were executed on a dedicated server equipped with an AMD Threadripper PRO 5955WX processor (16 cores, 32 threads, 3.8GHz), DDR4-3200 64GB ECC/REG × 8 (total 512GB RAM), and the Ubuntu 22.04 operating system.
For comparison, we evaluated CODA against the existing RR and MLS scheduling policies already available in the same simulation environment. All scheduling policies were tested on identical benchmark circuits with the same number of decoders, ensuring a fair and consistent comparison.
To ensure a reliable and reproducible evaluation, all quantum circuits used in this study were obtained from the publicly available MQT Bench suite~\cite{mqtbench}.
The performance of CODA was evaluated using two primary metrics. First, we measure the longest undecoded sequence length, which is defined as the maximum number of consecutive time slices during which a logical qubit remains undecoded, thereby reflecting the ability of the scheduling algorithm to suppress error accumulation. Second, we evaluate scalability by progressively increasing the circuit size and comparing the theoretical scheduling complexity bound with the actual simulated scheduling time.


\begin{figure*}[!t] 
    \centering
    \centering{\includegraphics[width=17.5cm]{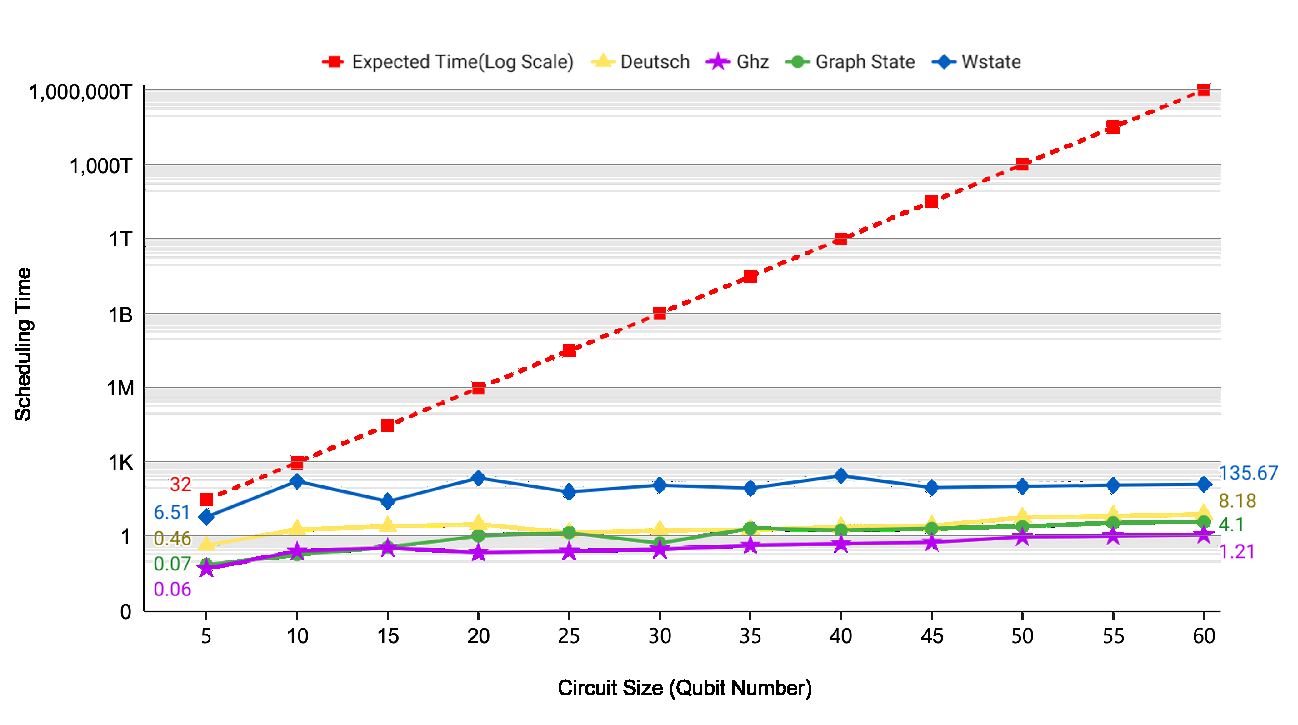}}
    \caption{The scalability of the CODA algorithm in terms of scheduling time. The benchmark circuits include Deutsch, GHZ, Graph State, and Wstate, all obtained from the publicly available MQT Bench~\cite{mqtbench}. The red line indicates the theoretical exponential upper bound (expected time) on a logarithmic scale, while the dotted lines show the CODA simulation results. The results demonstrate that, unlike the exponential growth of the theoretical bound, CODA exhibits only mild increases in runtime as circuit size grows, confirming its practicality as a scalable scheduling policy.}
    \label{fig:Scalability}
\end{figure*}

\subsection{Simulation Result}

Before discussing the benchmark results, it should be noted that both the decoder ratio ($N/M$) and the resulting feasible gap ($G$) are benchmark-dependent quantities in our evaluation. Specifically, for each benchmark, the number of decoders is determined according to the maximum number of simultaneous mandatory decoding demands associated with T-gate execution. Accordingly, the decoder ratio is not fixed across the benchmark suite, and the final feasible gap value also varies depending on circuit-specific factors such as qubit count, circuit depth, and the temporal distribution of T gates. For this reason, a single representative value of $G$ or $N/M$ is not sufficient to characterize all benchmarks.

\subsubsection{Longest Undecoded Sequence Lengths}

Figure~\ref{fig:Longest Undecoded Sequence Comparison} compares the longest undecoded sequence length achieved by three different scheduling policies: RR, MLS, and the proposed CODA across a variety of benchmark circuits. 
These 19 benchmarks were systematically selected to demonstrate the scheduling algorithms' performance comprehensively, reflecting two important considerations: (1) a thorough evaluation across different circuit scales, by evaluating on circuits with qubit counts ranging from relatively small (e.g., \texttt{qec-5}) to large (e.g., \texttt{adder-433}). (2) algorithm diversity, by incorporating diverse quantum algorithm types such as quantum walks (\texttt{qwalk-31}), machine learning (\texttt{dnn-51}), and physical simulation (\texttt{ising-66}).
The detailed numerical results for each benchmark are provided in Table~\ref{tab:benchmark_results}, specifically in the column denoting the longest undecoded sequence length.

The results reveal distinct performance differences among the three approaches. RR exhibits the lowest performance, showing the longest undecoded sequence lengths across all tested benchmarks among the three scheduling policies. The RR policy, like MLS and CODA, prioritizes qubits with imminent $T$-gates in each time slice; however, for the remaining qubits, decoders are assigned in a Round-Robin manner without accounting for the accumulation of undecoded sequences. 
As a result, certain qubits experience extended undecoded sequences, leading to noticeable error accumulation. 
Meanwhile, MLS mitigates these limitations to some extent by prioritizing qubits with the longest undecoded sequence at each time slice, thereby achieving better performance than RR. 
However, since MLS is a short-sighted policy that considers only the current slice, it fails to anticipate the mandatory decoding demands introduced by future $T$-gates. 
Consequently, when available decoders are exhausted by locally prioritized qubits, some qubits still experience unnecessarily long undecoded sequences.


By contrast, CODA consistently achieves the shortest undecoded sequence length across all benchmarks. 
While RR and MLS rely on heuristic scheduling strategies, CODA searches for a globally optimized schedule by incrementally increasing the longest undecoded sequence length limit $G$ from 1 and evaluating circuit-wide feasibility at each iteration. During this process, CODA simultaneously accounts for imminent $T$-gates and accumulated undecoded sequences in its scheduling decisions. 
This constraint-driven optimization strategy enables CODA to effectively suppress the growth of undecoded sequence length under limited decoder resources, thereby minimizing worst-case backlog accumulation. Quantitatively, in the small group, the \texttt{qec-5} benchmark shows that CODA reduces the longest undecoded sequence by 88.8\% (9$\to$1) compared to RR and by 87.5\% (8$\to$1) compared to MLS. 
In the medium group, \texttt{qwalk-31} demonstrates a 99.97\% reduction (3853$\to$1) relative to RR and an 85.7\% reduction (7$\to$1) relative to MLS. 
In the large group, \texttt{ghz-78} achieves a 92.8\% reduction (14$\to$1) compared to RR and an 87.5\% reduction (8$\to$1) compared to MLS. 
This overall superiority is visually captured by the g-mean bar, which confirms a significant average performance improvement across all tested benchmarks.
These results confirm that CODA consistently maintains the shortest undecoded sequence length regardless of circuit size, establishing it as the most effective scheduling policy for suppressing error accumulation under constrained decoder resources.

\subsubsection{Scalability}

The CODA algorithm addresses the fundamental intractability of decoder scheduling by transforming the exponentially complex global optimization problem into a sequence of time-bounded feasibility checks. Theoretically, finding a globally optimized schedule requires exploring a combinatorial space that expands according to the lower bound $|\Omega| \ge (\binom{N}{M})^{T}$, which grows super-exponentially with circuit size~\cite{garey1979computers}. Evaluating the feasibility of each mapping within this exploding space constitutes the dominant computational cost. To rigorously assess scalability against this theoretical hardness, we evaluated CODA on four benchmark circuits: Deutsch, GHZ, Graph State, and Wstate. Specifically, we utilized the publicly available MQT Bench suite~\cite{mqtbench} to generate instances of these circuits, varying the number of qubits to verify the algorithm's performance scaling.


Figure~\ref{fig:Scalability} illustrates the variation in scheduling time, contrasting the theoretical complexity with CODA's actual performance. The red dashed line, labeled as 'Expected Time (Log Scale)', represents the theoretical exponential upper bound. This exponential trajectory serves as a conservative baseline to demonstrate the prohibitive cost of exhaustive search strategies. It is derived from the worst-case scenario where the decoder resource ratio maximizes the search space (i.e., $M \approx N/2$). By applying \textit{Stirling's approximation}, we demonstrated that the combinatorial complexity in this worst-case asymptotically approaches $O(2^N)$~\cite{robbins1955remark}. A detailed mathematical analysis and the proof of this derivation are provided in Appendix~\ref{sec:Complexity}.

In stark contrast, the simulation results for CODA (solid lines) exhibit a marked deviation from this exponential curve. Instead of following the combinatorial explosion, the scheduling time adheres to a stable linear trajectory across all benchmark circuits. Specifically, as the circuit scale increased from 5 to 60 qubits, the runtime scaled linearly with the number of qubits, remaining within a practical range of 1.21 to 135.67 seconds. This linear scaling serves as empirical validation of our analysis in Appendix~\ref{sec:Complexity}, confirming that the time-bounded search strategy successfully decouples the scheduling cost from the exponential search space. Consequently, the runtime is constrained solely by the physical resource ratio ($\lceil N/M \rceil$), ensuring that CODA serves as a practical and scalable scheduling policy for future large-scale fault-tolerant quantum computing systems comprising hundreds of logical qubits.

\subsubsection{Other important metrics}

In this study, beyond the two primary metrics: the longest undecoded sequence length and scalability analysis, we additionally evaluated several secondary performance metrics to demonstrate the superiority of the proposed CODA over existing scheduling algorithms. 
These secondary metrics include decoder utilization, peak memory usage, and average undecoded sequence length. 
While the longest undecoded sequence length captures the worst-case peak, the average undecoded sequence length reveals whether the decoding workload is concentrated on specific qubits or evenly balanced across the system.
Decoder utilization represents the ratio of active decoders during the total scheduling period and serves as an indicator of how efficiently the limited decoding resources are employed. 
A lower utilization while maintaining comparable performance implies a more efficient use of decoders. 
Peak memory usage reflects the maximum volume of undecoded syndrome data stored in the buffer during execution~\cite{Johannes2024}. 
Detailed numerical results and benchmark-specific values for all these metrics are provided in Appendix~\ref{sec:Full Numerical Results}.

\section{\label{sec:Conclusion}Conclusion}
In this work, we addressed one of the key bottlenecks in fault-tolerant quantum computing, namely decoder scheduling, by proposing CODA, an optimization-based scheduling algorithm that outperforms the state-of-the-art heuristic MLS. 
We demonstrated the superiority of our approach through experimental results across 19 benchmark circuits. CODA achieved an average 74\% reduction in the longest undecoded sequence length compared to MLS, with a peak reduction of 85.7\% (from 7 to 1) observed in the \texttt{qwalk-31} circuit.
Crucially, regarding scalability, our evaluation confirms that CODA effectively bypasses the theoretical combinatorial explosion inherent in NP-hard scheduling problems. 
Instead of scaling exponentially, the scheduling runtime exhibits a robust linear trajectory governed by our constraint-based iterative search strategy, maintaining practical execution times (up to 135.67 seconds) even as circuits scale to 60 qubits.
These results demonstrate that CODA is a global optimization-based and highly scalable decoder scheduling solution, providing a practical foundation for efficient decoder virtualization in large-scale fault-tolerant quantum computing systems.

Building on this foundation, a promising direction for future work is the parallelization of scheduling computation across multiple candidate gap values. If supported by the underlying hardware, concurrent feasibility checks for different values of \(G\) could further reduce the wall-clock scheduling time of CODA without changing its global optimization framework. In particular, batched or coarse-grained exploration of candidate gap levels may provide a practical means to accelerate the search process in larger-scale settings.

\begin{acknowledgments}
This work was supported in part by the Institute for Information \& communications Technology Planning \& Evaluation (IITP) grant funded by the Korean government (MSIT) (No. RS-2020-II200014, A Technology Development of Quantum OS for Fault-tolerant Logical Qubit Computing Environment); the National Research Foundation of Korea (NRF) grant funded by the Korean government (MSIT) (No. RS-2025-25434537); and Creation of the quantum information science R\&D ecosystem (based on human resources) through the National Research Foundation of Korea (NRF) funded by the Korean government (Ministry of Science and ICT (MSIT)) (No. RS-2023-00256050).
\end{acknowledgments}

\appendix

\section{\label{sec:Scalability Limits of QEC Decoders in FTQC}Scalability Limits of QEC Decoders in FTQC}

\begin{figure}[!t]

    \centering{\includegraphics[width=0.455\textwidth]{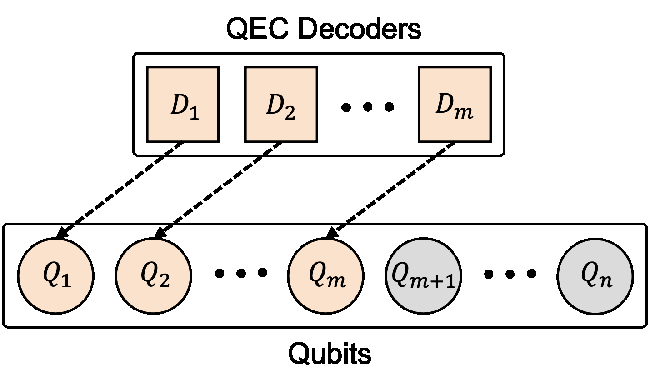}}
    \caption{Resource imbalance in large-scale FTQC system. Here, the scalability of decoder resources is limited by practical constraints such as hardware cost and power consumption, resulting in $m < n$ and creating a resource imbalance between available decoders and logical qubits.}
    \label{fig:FTQC resource}
\end{figure}

Large-scale FTQCs are projected to comprise millions of logical qubits. However, due to the physical and architectural constraints of cryogenic control systems, the number of decoders that can be integrated and powered within the system is expected to be limited to the order of hundreds to thousands~\cite{8901812}. As illustrated in Fig.~\ref{fig:FTQC resource}, this leads to a fundamental resource shortage ($m < n$), where $n$ logical qubits are serviced by only $m$ available decoders. Such a severe imbalance acts as a critical scalability limit, hindering the realization of large-scale FTQCs.

\begin{figure*}[!t]
    \centering{\includegraphics[width=17.5cm]{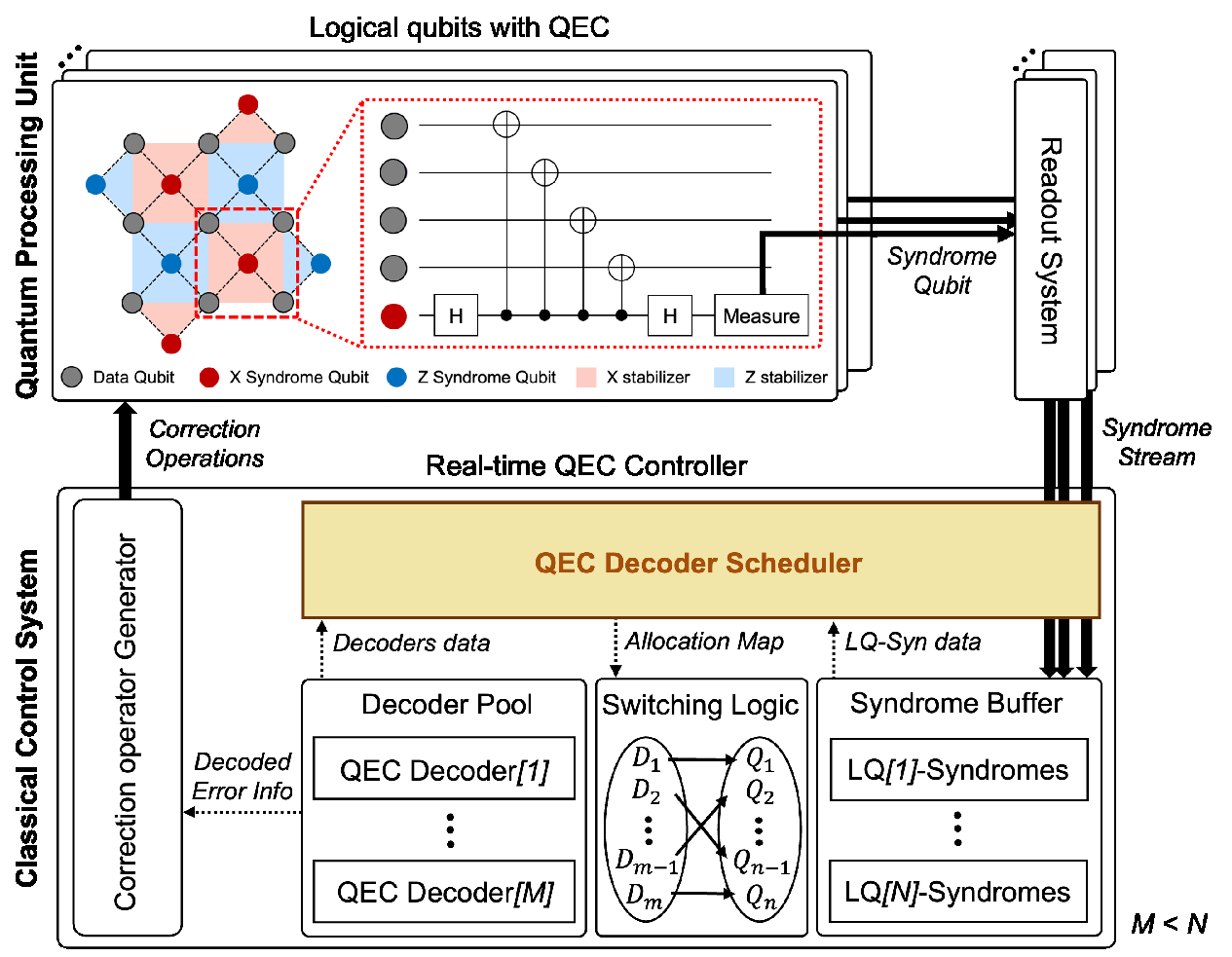}}
    \caption{This figure illustrates the system architecture of a Virtualized Quantum Decoder (VQD) environment. Syndrome information measured from the Quantum Processing Unit (QPU) is stored in the Syndrome Buffer via the Readout System. The QEC Decoder Scheduler generates an Allocation Map to assign a limited number of decoders to a larger number of logical qubits (LQs). The Switching Logic, implemented with a MUX/DEMUX structure, connects the selected LQ syndromes to the appropriate decoder. The decoded error information is then forwarded to the Correction Operator Generator, which produces the final correction operations. This architecture provides the baseline infrastructure for efficient decoder resource management and real-time error correction in the regime where $M < N$ (the number of decoders is smaller than the number of LQs).}
    \label{fig:QEC decoder overall process}
\end{figure*}

To illustrate the severity of this scalability challenge, consider the simplest hypothetical case where each logical qubit is paired with a dedicated decoder, a one-to-one allocation model. Under this baseline assumption, the system faces three significant constraints that fundamentally restrict its practical realization: (1) \textit{Area and Integration Limit}, (2) \textit{Power and Thermal Budget}, and (3) \textit{Data Routing Bottleneck}. These aspects are discussed in detail below.

\textit{Area and Integration Limit} 
QEC decoders are implemented as dedicated classical hardware such as FPGAs or ASICs, each requiring complex logic and substantial silicon area~\cite{liyanage2023fpga,bravyi_high-threshold_2024,nguyen2023qldpcfpga}. 
Although current decoders reside in the classical domain outside the cryogenic environment, future FTQC architectures are expected to demand tighter quantum–classical integration to reduce decoding latency~\cite{8901812,vandersypen_interfacing_2017}. 
Such proximity increases the density and complexity of interconnects within cryogenic control electronics, where limited wiring capacity and packaging space impose practical constraints on how many decoders can be co-located near the quantum hardware~\cite{charbon2021cryogenic,patra2023cryopackaging}. 
 
\textit{Power and Thermal Budget} 
Beyond the physical integration constraints, the scalability of decoder deployment is fundamentally limited by the restricted capacity of cryogenic refrigeration systems~\cite{PhysRevApplied.3.024010}. 
The available cooling power in such systems is only on the order of a few hundred milliwatts, most of which is already consumed by control and readout electronics~\cite{8901812}. 
As the number of decoders increases, the cumulative power dissipation can exceed this capacity, leading to local temperature rises that, in turn, degrade qubit coherence and reduce overall system stability~\cite{krinner_engineering_2019,wenner2013hotqp}.

\textit{Data Routing Bottleneck} 
In addition to integration and thermal constraints, the scalability of decoder deployment is also fundamentally limited by the bandwidth of data routing between the quantum processor and the classical decoding hardware. 
For a large-scale FTQC with millions of logical qubits, syndrome extraction typically occurs at MHz frequencies, generating on the order of tens to hundreds of bits per qubit per cycle~\cite{Gidney2021howtofactorbit}. 
This corresponds to aggregate data rates of tens of terabits per second, which far exceed the practical capacity of current cryo-to-room-temperature I/O interfaces~\cite{brennan2025classical}. 
This inherent routing bottleneck constitutes a fundamental scalability barrier, making massively parallel decoder architectures infeasible for future FTQC systems.

\section{\label{sec:Virtualization of Quantum Decoder Architecture}Virtualization of Quantum Decoder Architecture}

As large-scale FTQC systems face a structural resource imbalance where the number of available decoders ($m$) is significantly smaller than the number of logical qubits ($n$), the conventional one-to-one allocation model becomes impractical. This imbalance necessitates a new architectural paradigm to efficiently manage limited decoder resources. In recent studies, the concept of decoder virtualization has been proposed as a promising solution. This approach is analogous to time-multiplexing in classical computing systems~\cite{liu1973}, where a centralized pool of physical decoders is shared among multiple logical qubits across time slices, thereby enabling efficient utilization of limited resources.

The Virtualized Quantum Decoder (VQD) architecture systematizes this concept at the system level, as illustrated in Fig.~\ref{fig:QEC decoder overall process}. This figure depicts the overall system architecture where syndrome information measured from the Quantum Processing Unit (QPU) is first stored in the Syndrome Buffer via the Readout System. The QEC Decoder Scheduler generates an Allocation Map to assign limited decoders to logical qubits based on the current workload. Based on this map, the Switching Logic, typically implemented using MUX/DEMUX structures, connects the selected logical qubit syndromes to the appropriate physical decoder. Finally, the decoded error information is forwarded to the Correction Operator Generator, which produces the final correction operations.

Consequently, the performance of a decoder virtualization system fundamentally depends on the efficiency of its decoder scheduling policy, which determines which logical qubit should be allocated a limited decoder resource first. Maurya and Tannu introduced several heuristic-based static scheduling policies, including Most-Frequent Decoder (MFD), Round-Robin (RR), and Minimize Longest Undecoded Sequence (MLS)~\cite{maurya2024classicalqec}. Among them, MLS is an intuitive greedy algorithm that preferentially allocates an available decoder to the qubit that has accumulated the longest backlog of undecoded syndromes. While the VQD architecture and the MLS policy are pioneering works that demonstrated the feasibility of implementing FTQC under severe resource constraints, these approaches have fundamental limitations due to their reliance on local heuristics, which we address in this work.


\section{\label{sec:Related Works}Related Works}


In FTQC, QEC decoders are widely recognized as a key bottleneck for achieving temporal and spatial scalability. In large-scale systems, massive streams of syndrome data are generated at sub-microsecond timescales. Failing to process them in real time leads to the accumulation of undecoded syndromes, increased logical error rates, and excessive memory overhead. Consequently, recent research has focused on distinct layers of the stack, ranging from improving internal decoder architectures to optimizing system-level resource management.

Prior studies on large-scale simulation have recognized the necessity of sharing limited decoding resources across many logical qubits. XQsim~\cite{byun2022xqsim}, which models cross-technology control processors for 10K+ qubit quantum computers, explicitly adopts a Round-Robin (RR) scheduling policy to assign decoders to multiple logical qubits. However, it employs RR primarily as a baseline mechanism without deeply analyzing the efficacy or limitations of such scheduling policies. Similarly, AFS~\cite{das2022afs} focuses on accurate, fast, and scalable error decoding, highlighting the system-level need to manage decoders as a shared resource rather than dedicating one per logical qubit. While these works successfully identified the resource imbalance problem, they relied on simple heuristics like RR, leaving the potential of global optimization unexplored.


On the architectural front, LATTE~\cite{latte} serves as a representative example of efforts to enhance decoder throughput. LATTE proposes a dedicated decoding architecture that combines an FPGA-based neural local decoder with a CPU-based block decoder, streaming syndrome data across these components in both space and time. By carefully scheduling and distributing syndrome data to multiple decoding cores within a single decoder chip, LATTE achieves high throughput and low latency. Crucially, while LATTE primarily addresses intra-decoder dataflow and resource allocation (exploiting cores and memory inside a decoding unit), our work focuses on the system-level allocation problem, optimizing the distribution of limited decoder resources across multiple logical qubits.

\section{Problem Formulation, Optimization Strategy, and Scalability Analysis}
\subsection{\label{sec:Problem Formulation}Problem Formulation and Computational Hardness Analysis}

We formally define the decoder scheduling problem in resource-constrained FTQC systems as a global optimization task to determine the optimal allocation matrix over time $T = \{1, ..., L\}$, given a set of $N$ logical qubits $Q$ and $M$ decoders $D$ (where $M \ll N$). The fundamental computational difficulty of this problem stems from the recursive dependency of decisions along the temporal axis. Since any backlog from unallocated qubits at time $t$ carries over and accumulates at $t+1$, the global scheduling problem forms a single causal chain that cannot be decomposed into independent sub-problems.

Due to this state dependency, the size of the search space $\Omega$ exhibits exponential growth rather than linear scaling. Since the combinatorial choice of selecting $M$ decoders out of $N$ qubits accumulates over the circuit depth $L$, the mathematical lower bound of the search space size is derived as:
$$|\Omega| \ge \left( \binom{N}{M} \right)^L$$
This value explodes exponentially as the circuit scale increases, indicating that finding a mathematically perfect global optimum via exhaustive search belongs to the class of NP-hard problems.

To overcome this computational intractability, we adopt a strategy of transforming the original global optimization problem into a sequence of feasibility decision problems. Instead of directly minimizing the objective function, we reformulate the problem to determine whether a valid schedule exists under a given undecoded sequence length limit $G$ within a strict computation time limit. Consequently, the optimization goal is redefined as finding the minimum integer $G$ for which a feasible solution is found within the specified time limit. This formulation systematizes the uncertain search for an optimum into a time-bounded sequential verification process over the discrete parameter $G$. This serves as the theoretical basis for our proposed CODA algorithm, enabling the effective management of NP-hard complexity within practical computational limits.

\subsection{\label{sec:Constraints}Constraints}

To minimize the longest undecoded sequence length under limited decoder resources, CODA defines a concise set of structural constraints that capture the essential characteristics of the scheduling problem. These constraints determine the form of feasible allocations and define the search space of the gap-increment procedure.

We first define the main scheduling variables:

\begin{itemize}
    \item $x_{d,q,t} \in \{0,1\}$: Assignment variable. 
    $x_{d,q,t} = 1$ if decoder $d \in D$ is assigned to logical qubit $q \in Q$ at time slice $t \in T$, and $0$ otherwise.
    \item $a_{d,t} \in \{0,1\}$: Availability indicator. 
    $a_{d,t} = 1$ if decoder $d$ is available at time $t$.
    \item $y_{q,t} \in \{0,1\}$: Decoding indicator. 
    $y_{q,t} = 1$ if logical qubit $q$ is decoded by any decoder at time $t$, and $0$ otherwise, defined as:
    $$y_{q,t} = \sum_{d \in D} x_{d,q,t}.$$
    \item $U_q(t) \in \mathbb{Z}_{\ge 0}$: Backlog length. 
    The length of the undecoded syndrome sequence of logical qubit $q$ at time $t$.
    \item $G \in \mathbb{Z}_{\ge 1}$: Global gap limit. 
    The upper bound on the longest undecoded sequence length across all qubits and time slices.
    \item $\mathcal{T}_\tau \subseteq Q$: Set of qubits scheduled to execute a $T$ gate at time slice $\tau$. 
    These qubits receive higher priority in the scheduling process.
\end{itemize}

Time is discretized into slices $T = \{1, 2, \dots, L\}$. The optimization model enforces the following constraints to ensure physical feasibility and logical correctness

First, the model enforces resource constraints to adhere to physical hardware limits. Specifically, it guarantees that a decoder $d$ is assigned to at most one qubit only when available ($a_{d,t}=1$), and that no logical qubit is processed by multiple decoders simultaneously. These constraints are formulated as:
$$\sum_{q \in Q} x_{d,q,t} \le a_{d,t} \quad \forall d \in D, \forall t \in T$$

Second, the temporal backlog evolution of the backlog state $U_{q}(t)$ is governed by a recurrence relation. If a qubit is decoded ($y_{q,t}=1$), its backlog resets to zero; otherwise ($y_{q,t}=0$), it increments by one. Assuming an initial backlog $U_{q}(1)=0$, the relation is defined as:
$$U_{q}(t+1) = (1 - y_{q,t})(U_{q}(t) + 1) \quad \forall q \in Q, \forall t \in T \setminus \{L\}$$
When $y_{q,t}=1$, the right-hand side becomes zero, indicating that decoding occurred. When $y_{q,t}=0$, it becomes $U_{q}(t)+1$, reflecting the accumulation of latency.

Third, the algorithm enforces bound constraints such that the backlog of any qubit must not exceed the global limit $G$:
$$U_{q}(t) \le G \quad \forall q \in Q, \forall t \in T$$
In the iterative search procedure, this inequality constraint is progressively relaxed by incrementing $G$, thereby expanding the feasible search space until a valid allocation is found.

Finally, hard precedence constraints are applied to maintain the logical integrity of the quantum circuit. Any qubit scheduled for a T-gate at time $\tau$ must be decoded immediately prior to the gate execution. CODA imposes this as a hard constraint:
$$y_{q,\tau-1} = 1 \quad \forall \tau \in T \setminus \{1\}, \forall q \in \mathcal{T}_{\tau}$$
This strictly requires that every qubit in $\mathcal{T}_{\tau}$ completes decoding at slice $\tau-1$, preventing logical errors during non-Clifford operations.

\subsection{\label{sec:Optimization Strategy}Optimization Strategy}

The scheduling objective of CODA is to minimize the longest undecoded sequence length across all logical qubits and time slices:
\[
\min_{X} \left( \max_{q \in Q, t \in T} U_q(t) \right)
\]
where $X = \{x_{d,q,t}\}$ denotes the set of binary allocation variables over the entire circuit execution. This formulation addresses the minimization of the longest undecoded sequence length subject to the resource and precedence constraints described in Appendix~\ref{sec:Constraints}.

Instead of relying on a computationally expensive global optimization solver to minimize this objective directly, CODA employs an iterative feasibility search strategy. The CODA starts from an undecoded sequence length limit of $G = 1$ and incrementally increases $G$ while testing whether all constraints are satisfied. Once a feasible allocation is found, the search terminates, and the corresponding allocation is applied for scheduling. This approach identifies the minimum feasible undecoded sequence length $G$ that satisfies all constraints without performing an exhaustive combinatorial search over all possible allocations.

To ensure scalability, CODA incorporates a time-bounded search mechanism. If the feasibility check exceeds a predefined computation time limit, the algorithm immediately proceeds to the next parameter value $(G + 1)$ and repeats the process. This mechanism prevents the scheduling computation time from growing uncontrollably with system size and ensures predictable runtime even for large-scale FTQC workloads.

\subsection{\label{sec:Solver Configuration}Solver Configuration}

The CODA algorithm is implemented using the CP-SAT solver, integrating a gap-incremental search strategy with a time-bounded termination mechanism. This configuration maintains stable allocation quality and predictable computation time even under limited decoder resources. The key configuration parameters and the search procedure are as follows:

\begin{itemize}
    \item \textbf{Initial gap value} ($G_{\mathrm{init}}$):  
    The search begins at $G_{\mathrm{init}} = 1$, corresponding to the smallest possible longest undecoded sequence length. This establishes a bottom-up search trajectory, starting from the most constrained scenario.
    
    \item \textbf{Per-gap time limit} ($T_{\mathrm{limit}}$):  
    A strict computation-time budget, $T_{limit}$, is imposed for validating feasibility at each candidate gap $G$. If a feasible allocation is identified within this window, the search terminates immediately, and the corresponding solution is adopted as the final schedule.

    \item \textbf{Gap Increment and Relaxation}:  
    If no feasible allocation is found within $T_{limit}$ (due to timeout or infeasibility), the algorithm increments $G$ by one. This step constitutes a discrete relaxation of the constraints, expanding the search space before re-evaluating feasibility under the same time constraint. This process repeats until a valid solution is found.

    \item \textbf{Early Termination}:  
    The procedure halts immediately upon identifying the first feasible gap $G$. Since constraints are progressively relaxed as $G$ increases, the first feasible $G$ mathematically corresponds to the minimum feasible limit. Further exploration of larger $G$ values is unnecessary as it would yield suboptimal solutions.

\end{itemize}


This configuration provides a clear trade-off between allocation quality and computational efficiency. More precisely, \(T_{\mathrm{limit}}\) bounds only the computation time allocated to each gap evaluation, rather than terminating the entire scheduling procedure. If no feasible allocation is found within \(T_{\mathrm{limit}}\) for the current gap \(G\), due to timeout or infeasibility, CODA increments \(G\) by one, relaxes the constraints, and re-evaluates feasibility under the same time constraint. Accordingly, a smaller \(T_{\mathrm{limit}}\) enforces tighter scheduling overhead, but it may cause the algorithm to proceed to a larger gap before certifying feasibility at a smaller one. In contrast, a larger \(T_{\mathrm{limit}}\) gives the solver more opportunity to identify a feasible solution for smaller, more computationally difficult gap values, thereby improving the final allocation quality at the cost of longer computation time. In this sense, \(T_{\mathrm{limit}}\) directly influences scheduling performance by balancing scheduling overhead against the quality of the resulting allocation.

\subsection{~\label{sec:Complexity}Complexity and Scalability Analysis}

We analyze the theoretical hardness of the QEC decoder scheduling problem in large-scale systems and verify the scalability of the proposed CODA algorithm by examining its computational complexity and the scaling characteristics of the solution under physical resource constraints.

\noindent\textbf{A. Theoretical Hardness}

The computational difficulty of this problem stems from the recursive dependency where the allocation decision at the current time step dictates the system state for future steps. If limited decoder resources are assigned to specific qubits at time $t$, the processing requirements of the unassigned qubits do not disappear but are carried over to time $t+1$, determining the backlog load the system must handle. Since decisions at each moment are not independent but constrain future resource availability, the total number of scheduling scenarios accumulates multiplicatively at each step. As the circuit scale increases (with increasing qubit number $N$ and circuit depth $T$), the size of the total search space $|\Omega|$ expands exponentially according to the following lower bound:
$$|\Omega| \ge \left( \binom{N}{M} \right)^T$$
Due to this exponential expansion of the search space, finding a mathematically global optimum via exhaustive search is computationally intractable for large-scale circuits. This indicates that the problem belongs to the NP-hard class, justifying the necessity of a time-bounded approach to find a solution within a deterministic timeframe.

To rigorously evaluate the intractability depicted in Figure~\ref{fig:Scalability}, we further analyze the theoretical worst-case scenario derived from the lower bound. The search space is maximized when the decoder resource ratio creates the largest number of combinations (i.e., $M \approx N/2$). Applying \textit{Stirling's approximation} ($n! \approx \sqrt{2\pi n} (n/e)^n$), the asymptotic behavior of the combinatorial term in this worst-case is derived as equation~\ref{equ:Stirling's approximation}~\cite{robbins1955remark}.
\begin{equation}
\label{equ:Stirling's approximation}
\begin{aligned}
\binom{N}{N/2} &= \frac{N!}{((N/2)!)^2} \\
&\approx \frac{\sqrt{2\pi N} \cdot \frac{N^N}{e^N}}{\pi N \cdot \frac{(N/2)^N}{e^N}} \\
&= \sqrt{\frac{2}{\pi N}} \cdot \frac{N^N}{(N/2)^N} \\
&= \sqrt{\frac{2}{\pi N}} \cdot 2^N
\end{aligned}
\end{equation}
Although the term $\sqrt{\frac{2}{\pi N}}$ introduces a polynomial reduction, the growth rate is dominated by the exponential term $2^N$. Substituting this dominant term back into the total complexity equation, the full worst-case complexity scales super-exponentially as $O(2^{NT})$. However, visualizing this magnitude is impractical. Therefore, the theoretical exponential upper bound shown in Figure~\ref{fig:Scalability} plots the dominant exponential factor $2^N$. This serves as a conservative baseline to demonstrate the prohibitive cost of exhaustive search strategies.

\noindent\textbf{B. Computational Complexity of CODA}

To overcome this intractability, CODA employs a time-bounded gap search strategy. The algorithm iteratively performs feasibility checks by sequentially incrementing the backlog bound parameter $G$ from 1 until the first feasible solution is found at $G_{final}$. The total execution time is defined as the sum of the check durations for each step:
$$T_{total} = \sum_{G=1}^{G_{final}} T_{check}(G)$$
Since the check duration $T_{check}(G)$ for each step is strictly bounded by the user-defined parameter $T_{limit}$ ($T_{check} \le T_{limit}$), the upper bound of the total time complexity is derived as $O(G_{final} \cdot T_{limit})$. This implies that the computational complexity of CODA does not grow exponentially with circuit size but scales linearly with the solution quality metric $G_{final}$.

\noindent\textbf{C. Scalability and Convergence Analysis of CODA}

To demonstrate the practical scalability of CODA, we integrate the analysis of the theoretical lower bound of $G_{final}$ with the convergence characteristics derived from constraint relaxation. First, the scaling behavior of $G_{final}$ is fundamentally governed by physical resource constraints. Assuming a worst-case load scenario where all $N$ logical qubits require decoding at every time step, processing these requests with $M$ available decoders requires a minimum of $\lceil N/M \rceil$ time slots according to the Generalized Pigeonhole Principle~\cite{rosen_discrete_2019}. Consequently, the mathematical lower bound for the maximum backlog $G$ is established as:
$$G_{final} \ge \left\lceil \frac{N}{M} \right\rceil$$
This inequality proves that $G_{final}$ scales linearly ($\approx O(N)$) with the number of qubits $N$, rather than exponentially. This provides the theoretical basis for the runtime complexity of CODA, $O(G_{final} \cdot T_{limit})$, which maintains linearity with circuit size even in large-scale environments. Furthermore, the rapid convergence near this physical lower bound is mathematically guaranteed by the exponential expansion of the solution space due to constraint relaxation. Incrementing the search gap from $G$ to $G+1$ corresponds to a discrete relaxation that extends the valid time window $w_q$ for each qubit $q$. Since the set of globally feasible solutions $\mathcal{F}$ is formed by the Cartesian product of individual decision spaces, the expansion ratio of the feasible set scales according to the multiplication rule:
$$\frac{|\mathcal{F}(G+1)|}{|\mathcal{F}(G)|} \approx \prod_{q=1}^{N} \left( 1 + \frac{1}{w_q} \right)$$
Since every term $(1 + 1/w_q)$ is strictly greater than 1, the product indicates an exponential expansion of the feasible solution set with respect to the number of qubits $N$. As a result, the solution density within the search space rises drastically as $G$ increases. Therefore, the strategy of imposing a time limit ($T_{limit}$) effectively allows for a timeout at the computationally expensive tight bound (where solution density is critical) and induces a transition to the solution-rich relaxed region. This mechanism establishes a practical trade-off, sacrificing marginal optimality ($+1$ increase in backlog) in exchange for guaranteeing robust and fast convergence even in worst-case bottleneck scenarios.




\section{\label{sec:Full Numerical Results}Full Numerical Results}
Table~\ref{tab:benchmark_results} presents a comprehensive comparison across a variety of benchmark circuits of the performance of three scheduling policies: RR, MLS, and the proposed CODA. In addition to the longest undecoded sequence length, the table includes several supplementary metrics: decoder utilization (\textit{Decoder Utilization}), peak memory usage (\textit{Memory Usage}), the average length of undecoded sequences (\textit{Average Undecoded Sequence Length}), and the number of qubits exhibiting that average length (\textit{Q.Average Undecoded Sequence Length}). These metrics were selected as they provide critical means to directly evaluate the overall error-control capability of the system.

\textit{Decoder Utilization} indicates how much of the available decoder resources were actually used; a lower value implies that fewer decoders were sufficient to achieve effective error correction. The \textit{Average Undecoded Sequence Length} reflects the mean length of undecoded sequence length across qubits; smaller values imply that qubits are decoded more frequently, thereby suppressing error accumulation more effectively. \textit{Q.Average Undecoded Sequence Length} denotes the number of qubits that exhibit this average sequence length, where larger values imply that decoding resources are more evenly distributed across qubits, preventing error accumulation from being concentrated on a small subset. Finally, \textit{Memory Usage(MB)} quantifies the maximum accumulation of undecoded syndrome data in memory, with higher values indicating that qubits were left undecoded for longer periods.  

When evaluating the results across these metrics, CODA consistently outperformed the other policies. In terms of decoder utilization, CODA required a comparable or lower number of decoders than RR and MLS while achieving superior error suppression. For instance, in the small group benchmark \texttt{qec-5}, CODA reduced memory usage by approximately 15\% compared to MLS, despite using the same number of decoders. In the medium group benchmark \texttt{bv-30}, CODA reduced the average undecoded sequence length to 1, significantly shorter than RR (4.29) and MLS (3.48), while also increasing the number of qubits at this average by about 1.8$\times$ compared to MLS. In the large group benchmark \texttt{ising-66}, CODA reduced peak memory usage by more than 6\% compared to MLS, while simultaneously lowering the average undecoded sequence length by approximately 2.5$\times$.  

These results demonstrate that CODA is not narrowly optimized for specific qubits but instead provides balanced error control across the entire qubit set. By combining efficient resource utilization, reduced syndrome backlog, and equitable decoding distribution, CODA establishes itself as a more robust and reliable scheduling policy than existing approaches.

\begin{table*}[p!]
\centering
\caption{Quantum Benchmark Scheduling Results}
\label{tab:benchmark_results}
\renewcommand{\arraystretch}{1}
\scriptsize
\resizebox{\textwidth}{!}{%
\begin{tabular}{|c|c|c|c|c|c|c|c|c|}
\hline
\makecell{Circuit \\ Size} & \makecell{Quantum \\ Benchmark} & \makecell{Scheduling \\ Policy} &
\makecell{Longest\\Undecoded\\Sequence\\Length} &
\makecell{Used Decoder\\(per time slice)} &
\makecell{Decoder\\Utilization} &
\makecell{Memory\\Usage\\(MB)} &
\makecell{Average\\Undecoded\\Sequence\\Length} &
\makecell{Q.Average\\Undecoded\\Sequence\\Length} \\
\hline
\multirow{18}{*}{Small}
 & \multirow{3}{*}{qec-5} & RR   & 9    & 82.368 & 0.936 & 0.029  & 4.080  & 30 \\
 &                           & MLS  & 8    & 88.000      & 1.000      & 0.023  & 4.470  & 41 \\
 &                           & \cellcolor{gray!20}CODA & \cellcolor{gray!20}\textbf{1} & \cellcolor{gray!20}87.120 & \cellcolor{gray!20}0.990  & \cellcolor{gray!20}0.020 & \cellcolor{gray!20}1.000 & \cellcolor{gray!20}80 \\
\cline{2-9}
 & \multirow{3}{*}{lpn-5}  & RR   & 10   & 71.064 & 0.987 & 0.040  & 4.850  & 63 \\
 &                           & MLS  & 10   & 72.000      & 1.000      & 0.057  & 6.730  & 92 \\
 &                           & \cellcolor{gray!20}CODA & \cellcolor{gray!20}\textbf{3} & \cellcolor{gray!20}69.840 & \cellcolor{gray!20}0.970  & \cellcolor{gray!20}0.028 & \cellcolor{gray!20}3.000 & \cellcolor{gray!20}132 \\
\cline{2-9}
 & \multirow{3}{*}{qaoa-6} & RR   & 490  & 104.000 & 1.000      & 8.443  & 178.900 & 736 \\
 &                           & MLS  & 10   & 104.000      & 1.000      & 0.266  & 6.890  & 134 \\
 &                           & \cellcolor{gray!20}CODA & \cellcolor{gray!20}\textbf{3} & \cellcolor{gray!20}93.496 & \cellcolor{gray!20}0.899 & \cellcolor{gray!20}0.245 & \cellcolor{gray!20}3.000 & \cellcolor{gray!20}140 \\
\cline{2-9}
 & \multirow{3}{*}{vqe-8}  & RR   & 427  & 144.000 & 1.000      & 17.021 & 179.140 & 1024 \\
 &                           & MLS  & 11   & 144.000      & 1.000      & 0.602  & 7.600    & 158 \\
 &                           & \cellcolor{gray!20}CODA & \cellcolor{gray!20}\textbf{3} & \cellcolor{gray!20}142.560 & \cellcolor{gray!20}0.990  & \cellcolor{gray!20}0.534 & \cellcolor{gray!20}3.000 & \cellcolor{gray!20}216 \\
\cline{2-9}
 & \multirow{3}{*}{qpe-9}  & RR   & 1518 & 112.000 & 1.000      & 115.419 & 531.050 & 1840 \\
 &                           & MLS  & 7    & 112.000      & 1.000      & 1.133    & 5.210    & 98 \\
 &                           & \cellcolor{gray!20}CODA & \cellcolor{gray!20}\textbf{2} & \cellcolor{gray!20}87.808 & \cellcolor{gray!20}0.784 & \cellcolor{gray!20}1.112 & \cellcolor{gray!20}2.000 & \cellcolor{gray!20}100 \\
\cline{2-9}
 & \multirow{3}{*}{qft-10} & RR   & 3648 & 128.000 & 1.000      & 1571.533 & 1461.300 & 11169 \\
 &                           & MLS  & 9    & 128.000      & 1.000      & 1.757    & 6.520    & 135 \\
 &                           & \cellcolor{gray!20}CODA & \cellcolor{gray!20}\textbf{2} & \cellcolor{gray!20}122.624 & \cellcolor{gray!20}0.958 & \cellcolor{gray!20}1.710  & \cellcolor{gray!20}2.000 & \cellcolor{gray!20}140 \\
\hline
\multirow{21}{*}{Medium}
 & \multirow{3}{*}{bv-19}  & RR   & 47   & 95.808 & 0.998 & 0.228 & 9.710  & 115 \\
 &                           & MLS  & 13   & 96.000      & 1.000      & 0.194 & 9.280  & 136 \\
 &                           & \cellcolor{gray!20}CODA & \cellcolor{gray!20}\textbf{7} & \cellcolor{gray!20}90.336 & \cellcolor{gray!20}0.941 & \cellcolor{gray!20}0.169 & \cellcolor{gray!20}6.860 & \cellcolor{gray!20}233 \\
\cline{2-9}
 & \multirow{3}{*}{adder-28} & RR   & 67   & 71.568 & 0.994 & 0.238 & 27.210 & 140 \\
 &                             & MLS  & 6    & 72.000      & 1.000    & 0.046 & 4.080 & 43 \\
 &                             & \cellcolor{gray!20}CODA & \cellcolor{gray!20}\textbf{1} & \cellcolor{gray!20}52.128 & \cellcolor{gray!20}0.724 & \cellcolor{gray!20}0.042 & \cellcolor{gray!20}1.000 & \cellcolor{gray!20}44 \\
\cline{2-9}
 & \multirow{3}{*}{bv-30}  & RR   & 10   & 96.824 & 0.931 & 0.052 & 4.290  & 332 \\
 &                           & MLS  & 7    & 104.000    & 1.000      & 0.036 & 3.480  & 42 \\
 &                           & \cellcolor{gray!20}CODA & \cellcolor{gray!20}\textbf{1} & \cellcolor{gray!20}88.296 & \cellcolor{gray!20}0.849 & \cellcolor{gray!20}0.052 & \cellcolor{gray!20}1.000 & \cellcolor{gray!20}76 \\
\cline{2-9}
 & \multirow{3}{*}{qwalk-31} & RR   & 3853 & 120.000 & 1.000      & 1207.753 & 1169.990 & 2788 \\
 &                             & MLS  & 3    & 120.000    & 1.000      & 5.207    & 2.590    & 73 \\
 &                             & \cellcolor{gray!20}CODA & \cellcolor{gray!20}\textbf{1} & \cellcolor{gray!20}71.040 & \cellcolor{gray!20}0.592 & \cellcolor{gray!20}5.178 & \cellcolor{gray!20}1.000 & \cellcolor{gray!20}64 \\
\cline{2-9}
 & \multirow{3}{*}{gs-37}  & RR   & 68   & 103.688 & 0.997 & 0.295 & 9.420  & 37 \\
 &                           & MLS  & 10   & 104.000    & 1.000      & 0.208 & 7.330  & 119 \\
 &                           & \cellcolor{gray!20}CODA & \cellcolor{gray!20}\textbf{5} & \cellcolor{gray!20}101.712 & \cellcolor{gray!20}0.978 & \cellcolor{gray!20}0.257 & \cellcolor{gray!20}5.000 & \cellcolor{gray!20}223 \\
\cline{2-9}
 & \multirow{3}{*}{wstate-40} & RR   & 1350 & 120.000 & 1.000      & 111.633 & 525.250 & 1823 \\
 &                              & MLS  & 7    & 120.000    & 1.000      & 1.091    & 4.460    & 71 \\
 &                              & \cellcolor{gray!20}CODA & \cellcolor{gray!20}\textbf{1} & \cellcolor{gray!20}97.440 & \cellcolor{gray!20}0.812 & \cellcolor{gray!20}1.074 & \cellcolor{gray!20}1.000 & \cellcolor{gray!20}72 \\
\cline{2-9}
 & \multirow{3}{*}{multiplier-45} & RR   & 683  & 104.000 & 1.000      & 36.148 & 302.080 & 1128 \\
 &                                  & MLS  & 7    & 104.000    & 1.000      & 0.645  & 4.190    & 59 \\
 &                                  & \cellcolor{gray!20}CODA & \cellcolor{gray!20}\textbf{1} & \cellcolor{gray!20}70.616 & \cellcolor{gray!20}0.679 & \cellcolor{gray!20}0.635 & \cellcolor{gray!20}1.000 & \cellcolor{gray!20}60 \\
\hline
\multirow{18}{*}{Large}
 & \multirow{3}{*}{dnn-51} & RR   & 4527 & 128.000 & 1.000      & 1316.260 & 1414.040 & 5310 \\
 &                           & MLS  & 7    & 128.000    & 1.000      & 4.965    & 5.240    & 111 \\
 &                           & \cellcolor{gray!20}CODA & \cellcolor{gray!20}\textbf{2} & \cellcolor{gray!20}94.208 & \cellcolor{gray!20}0.736 & \cellcolor{gray!20}4.534 & \cellcolor{gray!20}2.000 & \cellcolor{gray!20}112 \\
\cline{2-9}
 & \multirow{3}{*}{wstate-60} & RR   & 1834 & 120.000 & 1.000      & 393.154 & 805.320 & 2738 \\
 &                              & MLS  & 7    & 120.000    & 1.000      & 2.509    & 4.610  & 75 \\
 &                              & \cellcolor{gray!20}CODA & \cellcolor{gray!20}\textbf{1} & \cellcolor{gray!20}102.720 & \cellcolor{gray!20}0.856 & \cellcolor{gray!20}2.478 & \cellcolor{gray!20}1.000 & \cellcolor{gray!20}60 \\
\cline{2-9}
 & \multirow{3}{*}{deutsch-60} & RR   & 78   & 103.480 & 0.995 & 0.967 & 27.580 & 190 \\
 &                               & MLS  & 7    & 104.000    & 1.000      & 0.175 & 4.540  & 75 \\
 &                               & \cellcolor{gray!20}CODA & \cellcolor{gray!20}\textbf{3} & \cellcolor{gray!20}97.032 & \cellcolor{gray!20}0.933 & \cellcolor{gray!20}0.157 & \cellcolor{gray!20}3.000 & \cellcolor{gray!20}148 \\
\cline{2-9}
 & \multirow{3}{*}{ising-66} & RR   & 1328 & 272.000 & 1.000      & 177.629 & 499.790 & 4139 \\
 &                            & MLS  & 16   & 272.000    & 1.000      & 2.462    & 10.250  & 303 \\
 &                            & \cellcolor{gray!20}CODA & \cellcolor{gray!20}\textbf{4} & \cellcolor{gray!20}250.240 & \cellcolor{gray!20}0.920  & \cellcolor{gray!20}2.322 & \cellcolor{gray!20}4.000 & \cellcolor{gray!20}516 \\
\cline{2-9}
 & \multirow{3}{*}{ghz-78}  & RR   & 14   & 112.320 & 0.936 & 0.153 & 5.580 & 81 \\
 &                           & MLS  & 8    & 120.000    & 1.000      & 0.103  & 4.910  & 79 \\
 &                           & \cellcolor{gray!20}CODA & \cellcolor{gray!20}\textbf{1} & \cellcolor{gray!20}92.640 & \cellcolor{gray!20}0.772 & \cellcolor{gray!20}0.087 & \cellcolor{gray!20}1.000 & \cellcolor{gray!20}80 \\
\cline{2-9}
 & \multirow{3}{*}{adder-433} & RR   & 635  & 248.000 & 1.000      & 131.544 & 274.670 & 1708 \\
 &                              & MLS  & 4    & 248.000    & 1.000      & 2.741    & 3.070    & 59 \\
 &                              & \cellcolor{gray!20}CODA & \cellcolor{gray!20}\textbf{1} & \cellcolor{gray!20}209.808 & \cellcolor{gray!20}0.846 & \cellcolor{gray!20}2.626 & \cellcolor{gray!20}1.000 & \cellcolor{gray!20}172 \\
\hline
\end{tabular}%
}
\end{table*}

\clearpage

%

\end{document}